\documentclass[9pt,twocolumn,twoside]{pnas-new}

\templatetype{pnasresearcharticle} 
\title{A deep Aurum reservoir: Stable compounds of two bulk-immiscible metals under pressure}

\author[a,b,e]{Adebayo A. Adeleke}
\author[b,1]{Stanimir A. Bonev} 
\author[b]{Christine J. Wu}
\author[c,d]{Ericmoore E. Jossou}
\author[e,1]{Erin R. Johnson}

\affil[a]{Department of Physics and Engineering Physics, University of Saskatchewan, Saskatoon, Saskatchewan S7N 5E2, Canada}
\affil[b]{Quantum Simulation Group, Materials Science Division, Lawrence Livermore National Laboratory, Livermore, California 94550, USA}
\affil[c]{Department of Mechanical Engineering, University of Saskatchewan, Saskatoon, Saskatchewan S7N 5A9, Canada}
\affil[d]{Nuclear Science \& Technology Department, Brookhaven National Laboratory, Upton, New York 11973, USA}
\affil[e]{Department of Chemistry, Dalhousie University, Halifax, Nova Scotia B3H 4R2, Canada}

\leadauthor{Adeleke} 

\significancestatement{We have demonstrated that gold (Au), a noble and precious metal, can react with hexagonal close packed iron (Fe) to form stable intermetallic compounds under sufficient compression. We also demonstrated that Au, which is diamagnetic at ambient pressure, could attain a magnetic moment in compound form when compressed to the pressures found in the Earth's interior. The Fe-Au compounds show significant lattice thermal conductivities and their sound velocities have better agreement with seismic data relative to some other available binary models of the Earth's core. Our results suggest that the Earth's core could hold more Au than previously thought.}

\authorcontributions{S.A.B. and  A.A.A. designed research; A.A.A. performed research; E.E.J. and A.A.A. performed thermal conductivity calculations; A.A.A., S.A.B., C.J.W., E.E.J. and E.R.J. analyzed data; A.A.A. wrote the initial draft of the paper; A.A.A., S.A.B. and E.R.J. wrote the advanced version of the paper with contribution from all authors; S.A.B., C.J.W. and E.R.J. contributed funding and computer resources; S.A.B., C.J.W. and E.R.J. supervised research.}
\authordeclaration{The authors declare no conflicting interest.}
\correspondingauthor{\textsuperscript{1}To whom correspondence should be addressed. E-mail: S.A.B.:  bonev2@llnl.gov; E.R.J.: erin.johnson@dal.ca}

\keywords{Gold $|$ Intermetallic compound $|$ Earth's interior $|$ Iron} 

\begin{abstract}
The Earth's crust is known to be depleted of gold, among other slightly 
heavy noble metals transported by magma from the Earth's mantle to the 
crust. The bulk silicate Earth (BSE) model also suggests significant 
depletion of Au in the silicate mantle itself, which cannot be explained 
by the amount of Au in the mantle's magma. This implies that Au could 
remain in the lower mantle and form stable compounds, especially with 
iron, which is the predominant element within the core. While Fe does 
not form binary compounds or a bulk alloy with Au under ambient 
conditions, it may do so at the elevated pressures found in the Earth's 
interior. Here, using density-functional methods, we investigated the 
possibility of identifying stable, binary Fe-Au compounds at pressures
up to 210~GPa. We found three such Fe-Au compounds, which are stabilized 
by pressure and notable electron transfer, including an orthorhombic 
AuFe$_4$ phase that is ferromagnetic in nature with Au possessing a 
significant magnetic moment. While our results suggest that thermal 
convection due to the conductivities and the heat flux from the Fe-Au 
compounds could be an energy source to power the Earth's geodynamo, they 
also point towards changes in Au's chemical properties, as it can exist 
as either an anion or cation under pressure. In addition, the sound 
velocity and the density predicted for the various Fe-Au compounds 
suggest that they could shed light on the composition of the core-mantle 
boundary and the Earth's core, while demonstrating how the presence of 
trace amounts of Au could influence agreement with seismic data.

\end{abstract}

\dates{This manuscript was compiled on \today}
\doi{\url{www.pnas.org/cgi/doi/10.1073/pnas.XXXXXXXXXX}}

\begin{document}

\maketitle
\thispagestyle{firststyle}
\ifthenelse{\boolean{shortarticle}}{\ifthenelse{\boolean{singlecolumn}}{\abscontentformatted}{\abscontent}}{}



\dropcap{S}egregation of the proto-Earth's iron-rich core from its 
surrounding silicate, and collision-induced aggregation of 
planetesimals, resulted in the formation of the Earth within a period 
of 30-40 million years \cite{wood2006accretion}. Geological and 
seismological models for various regions of the Earth are constrained 
through comparisons with the composition of meteorites, which are 
similar to the planetesimals that helped form our planet 
\cite{stevenson1981models}. These direct comparisons have led to the 
conclusion that the Earth's crust was depleted of some siderophile 
elements, such as nickel, platinum, and gold. Lithophiles (with 
pressure-induced siderophilic properties), such as vanadium, chromium, 
manganese, and cesium, may have also percolated into the silicate mantle 
and the Earth's core \cite{jones1968gold,mcdonough1995composition}. 
Thus, the composition of the silicate mantle is expanded to include Cs, Ni,
Pt, Au, V, Cr and Mn \cite{ahrens1995global, 
helffrich2001earth, mcdonough1995composition}. 

Of these elements, Au is of particular 
interest because of its economic value \cite{mckay2017midas}, 
inertness \cite{briggs2019measurement}, and the remarkable stability 
of its crystal structure over a wide range of pressures traversing 
that of the Earth's core \cite{dubrovinsky2007noblest,
ishikawa2013pressure,ahuja2001theoretical,soderlind2002comment,
boettger2003theoretical}.  
In addition to depletion of Au from the Earth's crust, the bulk silicate 
Earth (BSE) model also suggests significant depletion of Au from the 
silicate mantle, which cannot be explained by the amount of Au in the 
mantle's magma \cite{wood2006accretion}. Since Au is heavy and less 
volatile than Cr, V, and Mn \cite{wasson1985meteorites,
wood2006accretion}, it could not have evaporated into space.
However, it is challenging to propose a likely form for Au in the 
neighborhood of the core-mantle boundary (CMB) without violating 
various constraints imposed in seismological models 
\cite{dziewonski1981preliminary,tkalvcic2018shear} of the Earth. 
A possible solution was proposed by Wood and coworkers 
\cite{norris2017earth}, who suggested that the Au further percolated 
into the Earth's iron-rich core.

Meanwhile, seismological studies \cite{dewaele2006quasihydrostatic} 
indicate a 2-5\% density deficit for the Earth's solid inner core, 
proposing that the core is not a `bank' of pure Fe  after all. 
The core must, therefore, be home to some light siderophile 
elements, such as H, C, Si, S, and K \cite{allegre1995chemical, 
adeleke2020formation,adeleke2020two,poirier1994light}. Additionally,
experiments at elevated temperatures and pressures, combined with 
theoretical studies, suggest the reactivity of the noble gases Xe 
\cite{stavrou2018synthesis} and Ar \cite{adeleke2019high} with Ni and 
Fe. The stability of the resulting compounds was attributed to 
pressure-induced energy raising of the valence p-shell states of the 
noble gas, allowing charge transfer to the partially filled 3d or 4s 
states of the transition metal \cite{stavrou2018synthesis}. However, 
none of these studies could exhaustively explain the mass and density 
deficit in the Earth's core from the preliminary reference Earth model 
(PREM) \cite{dziewonski1981preliminary}. This means that slightly heavy 
(siderophile) metals depleted in the Earth's crust could have found 
their way into the core, where they form stable compounds
 \cite{jones1968gold}. Thus, knowing the form(s) in which Au could 
exist when subject to the thermodynamic conditions present within the 
core may help provide a solution to the problem of depleted Au.  

It is well known that the Earth's core is predominantly composed of Fe 
\cite{allegre1995chemical,dziewonski1981preliminary,
mcdonough1995composition}. Thus, a natural line of thought would be that 
Au, which was drawn towards and into the core, may have formed a 
bimetallic compound with Fe. However, under ambient conditions, the 
magnetic metals (Fe, Co, and Ni) do not form compounds or bulk alloys 
with many heavier metals, including Au. This is primarily because 
of the size mismatch between the constituent atoms, coupled with little 
or no solid solubility \cite{nielsen1993initial,mehendale2010ordered}. 
If a stable, binary Fe-Au compound is demonstrated to form under high 
pressures, transversing that of the Earth's interior, we can then begin 
to search for Au in this form within the mantle and core.

To this end, we explore the Fe-Au potential energy surface from ambient 
to high pressure using density-functional theory (DFT). Interesting 
physics plays out under pressure and, as such, ushers in new chemistry 
\cite{dong2015chemical}. This is exemplified by the predicted formation 
of stable intermetallic compounds of Fe and Au under high pressure, 
reported here for the first time. Analysis of the electron density 
\cite{tang2009grid} reveals an unusual charge transfer between Au and 
Fe. Using a combination of DFT and the Boltzmann transport equation, we 
also predict the phonon-assisted thermal conductivity of the Fe-Au 
compounds at thermodynamic conditions relevant to the lower mantle 
(near the CMB) and the outer core. This work contributes to our 
understanding of the Earth's lower mantle and the outer core, revealing how it is able to keep such a noble metal as Au mixed with Fe without decomposition into its
elemental form. 
 
\section*{RESULTS}

\subsection*{Phase stability and stable crystal structures of Fe-Au}

We systematically searched for stable structures of the binary compounds 
Fe$_{x}$Au$_{y}$ ($x,y\in \{1\dots 4\}$) with cells containing 1-4 
structural formula units at 0~K temperature and pressures ranging from
0-200~GPa. Spin-polarized calculations were carried out on the 
lowest-energy structures for all stoichiometries explored and the 
nonmagnetic configurations were found to be the most energetically 
favorable,  with an exception observed for the AuFe$_{4}$ stoichiometry at 140~GPa.

\begin{figure}[h]
	\centering
	\includegraphics[width=1.0\linewidth]{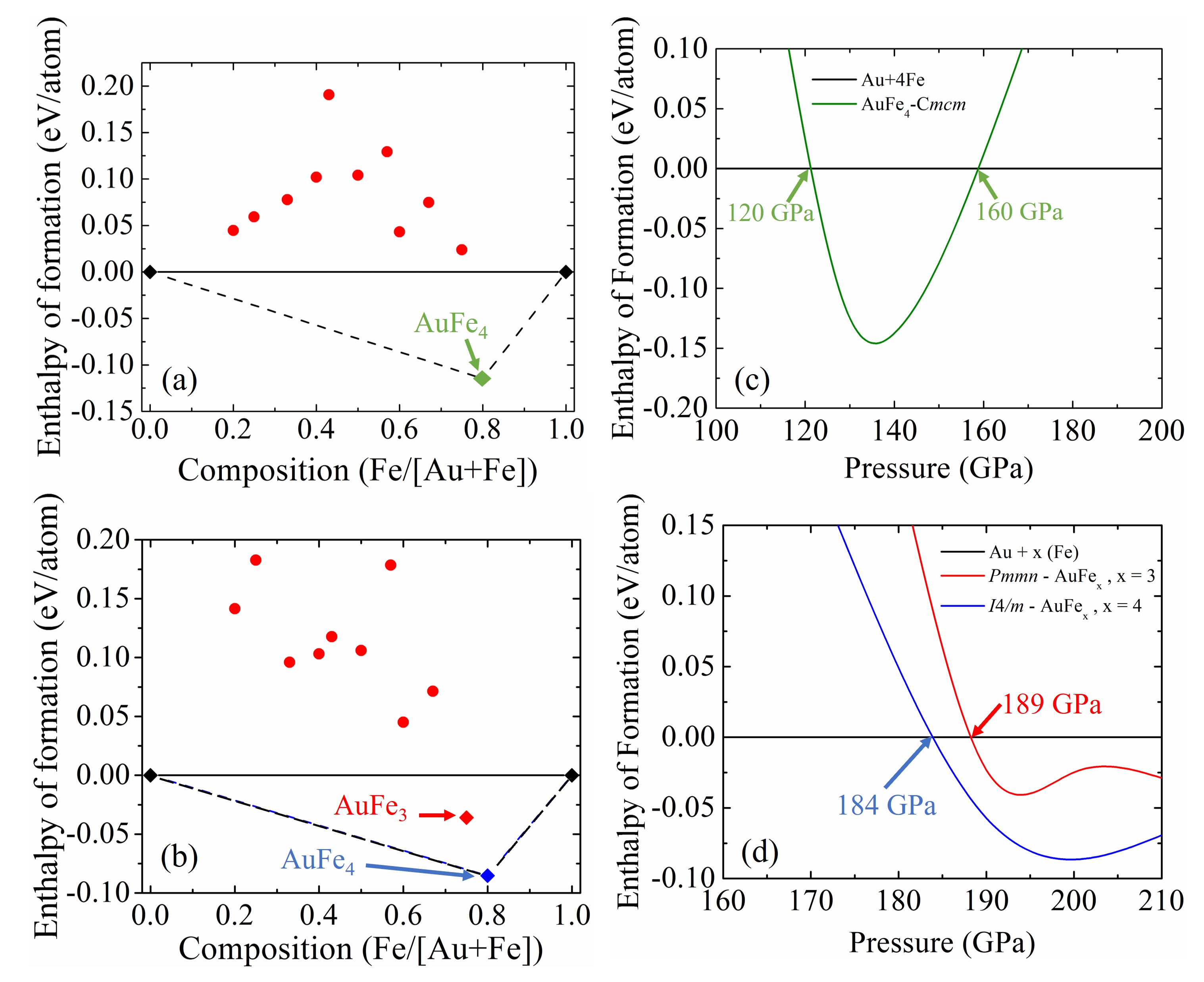}
	\caption{(color online) Calculated cold enthalpy (U+pV) of formation (0 K)  of various Fe-Au compounds with respect to constituent elemental decomposition (a) at 140 GPa (b) at 200 GPa. Calculated enthalpies per atom for (c) Cmcm-AuFe$_{4}$ structure with respect to the mixture of elemental Fe and Au (d) Pmmn-AuFe$_{3}$ and I4/m-AuFe$_{4}$ structures with respect to the mixture of elemental Fe and Au.}
	\label{fig:EOS}
\end{figure}

\begin{table*}
\centering
\caption{Structural parameters (and Bader atomic charges) of various 
predicted Fe-Au phases at 0 K}
\begin{tabular}{lcclcccr}
	\hline
	System & P (GPa) & SG & Lattice parameter & Element & Wyc. site & Atomic coordinate (fractional) & Bader charge/atom \\
	\midrule
	AuFe$_{4}$ & 140 & Cmcm & a = 12.12 Å, b = 4.00 Å, c = 3.69 Å & Au & 4c & 0.000, 0.796, 0.750 & 0.34 \\
	& & & & Fe & 8g & 0.883, 0.325, 0.750 & 0.33 / 0.37 \\
	& & & & Fe &8g & 0.705, 0.333, 0.750 & -0.51 / -0.53 \\
	\midrule
	AuFe$_{4}$ & 200 & I4/m & a = 10.00 Å, c = 10.01 Å & Au & 2b & 0.000, 0.000, 0.500 & 0.40\\
	& & & & Fe & 8h & 0.091, 0.690, 0.000 & -0.10\\
	\midrule
	AuFe$_{3}$ & 200 & Pmmn & a = 2.35 Å, b = 2.95 Å, c = 9.86 Å & Au & 2b & 0.000, 0.500, 0.585 & -0.26  \\
	& & & & Fe & 2b & 0.000, 0.500, 0.833 & 0.04\\
	& & & & Fe & 2b & 0.500, 0.000, 0.947 & 0.03 / 0.04\\
	& & & & Fe & 2b & 0.500, 0.000, 0.729 & 0.18 / 0.19\\
	\bottomrule
\end{tabular}
\label{table:structure}
\end{table*}

The convex hull calculated at various search pressures shows 
that Au and Fe do not form any stable binary compounds at ambient 
conditions up to \ 140~GPa (Fig. \ref{fig:EOS}). At this pressure, 
an orthorhombic AuFe$_{4}$ phase (with space group Cmcm) becomes 
thermodynamically stable relative to elemental fcc-Au and hcp-Fe 
(see Fig.~\ref{fig:EOS}a) and, as such, should be synthesizable.
Furthermore, at 200 GPa, AuFe$_{4}$ crystallizes into a tetragonal cell 
with space group I4/m. An orthorhombic AuFe$_{3}$ phase (space group 
Pmmn) that is metastable with respect to the tetragonal I4/m-AuFe$_{4}$ 
was also uncovered (see Fig.~\ref{fig:EOS}b). Inclusion of the 
free-energy contribution from lattice vibrations at 0~K (zero-point 
energy, ZPE) did not change the formation pressure, destabilize 
stable phases, nor stabilize other phases that were not initially 
energetically favorable. However, the magnitude of the formation 
enthalpy (enthalpy being U+pV) for the stable phases were slightly changed (Fig. S1).

The calculated equations of states (EOS) for the predicted structures 
reveal that the Cmcm-AuFe$_{4}$ has a stability pressure range of 
120-160~GPa (Fig.~\ref{fig:EOS}c) and, as such, will decompose below 
120~GPa and above 160~GPa. This explains why our structure searches at 
100~GPa and 200~GPa could not find this structure. On the other hand, 
the I4/m-AuFe$_{4}$ and Pmmn-AuFe$_{3}$ phases are thermodynamically 
stable from 184 and 189~GPa, respectively, up to at least 210~GPa (Fig. \ref{fig:EOS}d). 

\subsection*{Structural geometry of Fe-Au phases}

\begin{figure}
\centering
\setlength{\tabcolsep}{2pt}
\begin{tabular}{rlrlrlrl}
\raisebox{0.19\textheight}{a)} & \includegraphics[height=0.2\textheight]{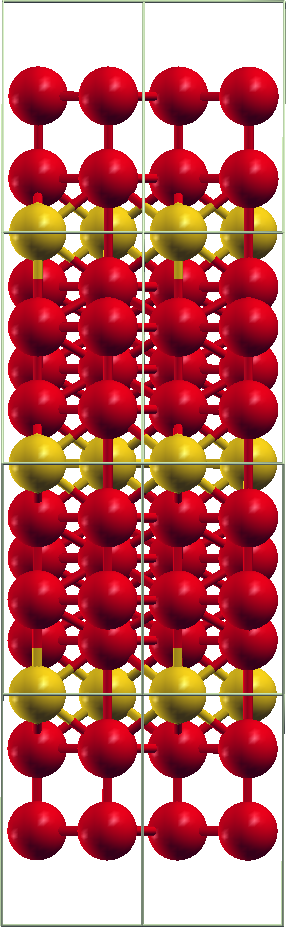} & 
\raisebox{0.19\textheight}{b)} & \includegraphics[height=0.2\textheight]{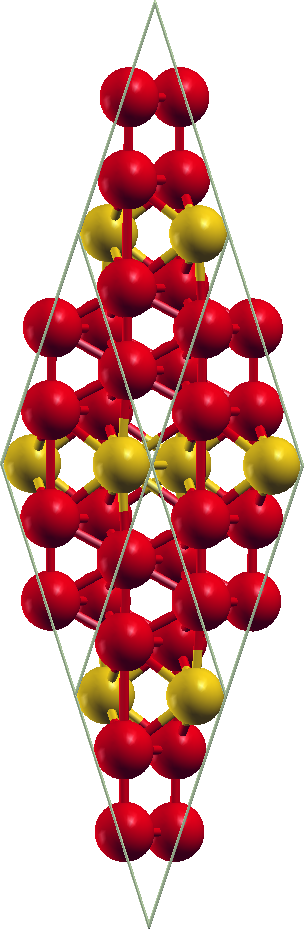} &
\raisebox{0.19\textheight}{c)} & \raisebox{0.02\textheight}{\includegraphics[height=0.16\textheight]{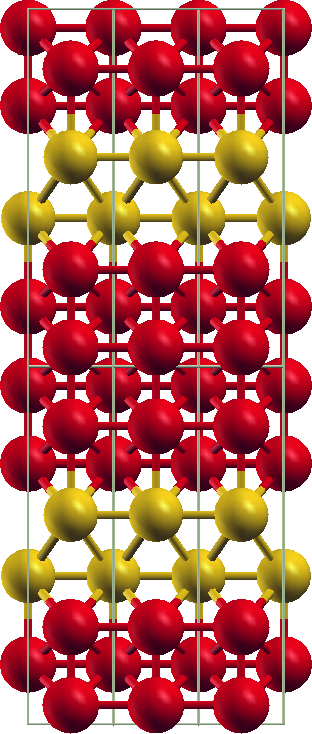}} &
\raisebox{0.19\textheight}{d)} & \raisebox{0.02\textheight}{\includegraphics[height=0.16\textheight]{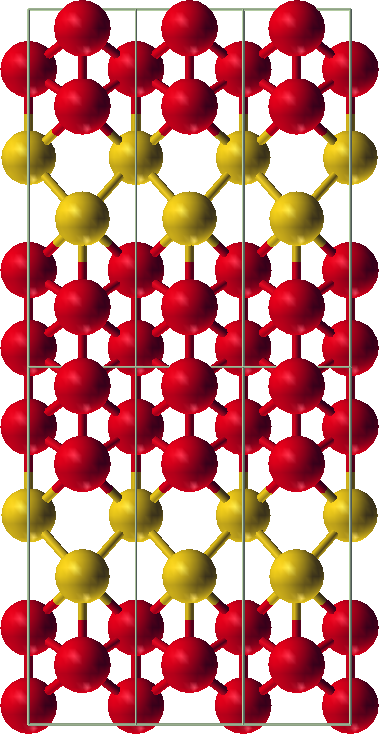}} \\
\end{tabular}

\vspace{\baselineskip}

\begin{tabular}{rlccrl}
\raisebox{0.09\textheight}{e)} & \includegraphics[height=0.1\textheight]{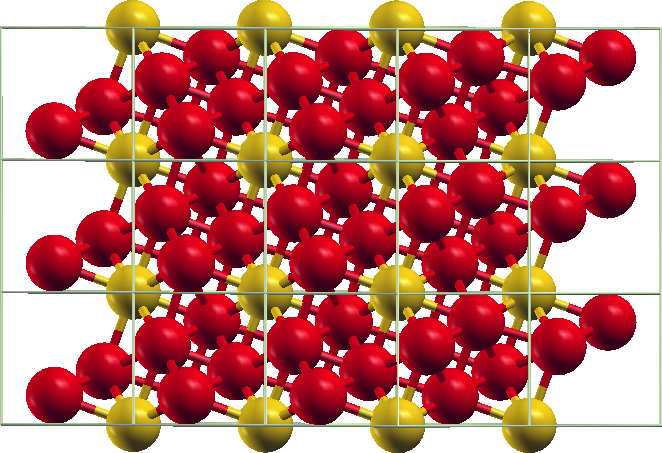} &  &  &
\raisebox{0.09\textheight}{f)} & \includegraphics[height=0.1\textheight]{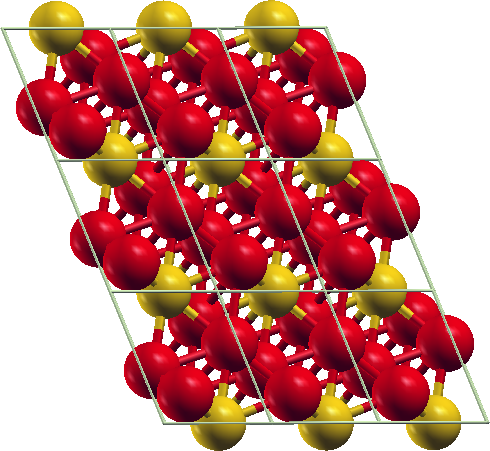} \\
\end{tabular}

\caption{(color online) Crystal structures of Cmcm-AuFe4 at 140~GPa (a,b),
Pmmn-AuFe3 at 200~GPa (c,d), and I4/m-AuFe4 at 200~GPa (e,f). Two views 
are shown for each crystal.}
\label{fig:Structure}
\end{figure}

The structural parameters of the thermodynamically stable phases of 
the Fe$_x$Au$_y$ compounds are presented in Table~\ref{table:structure} 
and the unit-cell geometries are shown in Figure~\ref{fig:Structure}.
The Cmcm-AuFe$_{4}$ phase at 140~GPa is characterized by layers of 
distorted hexagonal-close-packed Fe atoms connected by a layer of
Au atoms (Fig.~\ref{fig:Structure}a,b). This leads to two distinct Fe 
environments -- those closest to the Au and those in the interior of 
the Fe layer. As we will see, the presence of two distinct Fe environments 
facilitates charge transfer. The Fe-Au bond lengths range from 2.34-2.53 
\r{A}, the Au-Au bonds are 2.45 \r{A}, and the Fe-Fe bonds range from 
2.16-2.30 \r{A} (compared to 2.21 \r{A} for hcp-Fe at 140 GPa). 

Metastable Pmmn-AuFe$_{3}$ at 200~GPa has a phase-separated, layered 
structure. Six rows of Fe atoms form a hexagonal-close-packed 
arrangement, separated by a double layer of Au atoms
(Fig.~\ref{fig:Structure}c,d) stacked along the c direction. This 
leads to three distinct Fe environments that vary by their proximity to 
the Au layer. The Fe-Au bond lengths are 2.36 and 2.47 \r{A}, while
the Au-Au bond lengths are 2.35 \r{A} and the Fe-Fe bond lengths range
from 2.15-2.35 \r{A}.

In the I4/m-AuFe$_{4}$ structure at 200~GPa, Au atoms adopt 
body-centred-cubic positions, forming bonds with 12 symmetry-equivalent 
Fe atoms at the midpoints of each side of the surrounding cube 
(Fig.~\ref{fig:Structure}e). The Fe-Au bond length is 2.29 \r{A}, which 
is shorter than that observed in the other two crystals, and therefore 
stronger if we follow the `shorter = stronger' maxim for chemical bonds. 
The Fe-Fe bond lengths range from 2.14-2.35 \r{A} and there are no 
Au-Au bonds in this structure.

\subsection*{Magnetic properties of Fe-Au phases}

We performed spin-polarized calculations on the Cmcm-AuFe$_{4}$ 
phase at 140 GPa, and on $2 \times 2 \times 2$ supercells of the 
Pmmn-AuFe$_{3}$ and I4/m-AuFe$_{4}$ phases at 200 GPa. The results 
reveal that the ferromagnetic configuration, with a magnetic moment of 
$2.15\mu_{B}$, is the ground state for Cmcm-AuFe$_{4}$. This 
ferromagnetic configuration is 0.4 meV/atom lower in energy than the 
nonmagnetic configuration and $\sim$0.17 eV/atom lower in energy than 
the paramagnetic, antiferromagnetic, and ferrimagnetic configurations. 
While elemental gold exists as a diamagnet in bulk form 
\cite{jansen2008chemistry}, our results indicate that gold can adopt 
an induced ferromagnetic moment (of $0.52\mu_{B}$ in this case) when 
placed in the ferromagnetic environment of the surrounding iron. This 
is consistent with previous findings for other materials 
\cite{shih1931magnetic, kaufmann1945magnetization, 
pan1942ferromagnetic, stamatelatos2019paramagnetic}.

Conversely, the nonmagnetic configuration is preferred for both 
Pmmn-AuFe$_{3}$ and I4/m-AuFe$_{4}$ at 200~GPa. For both of these 
phases, the formation energy is ~1.2 meV/atom lower than the 
ferromagnetic case and ~0.2 eV/atom lower than the paramagnetic, 
antiferromagnetic, and ferrimagnetic configurations. This is not 
surprising since 200~GPa already exceeds the magnetic transition 
pressure for most transition metals \cite{zheng1998comment}.

\subsection*{Electronic structure of Fe-Au phases}

\begin{figure}
	\centering
	\includegraphics[width=1.0\linewidth]{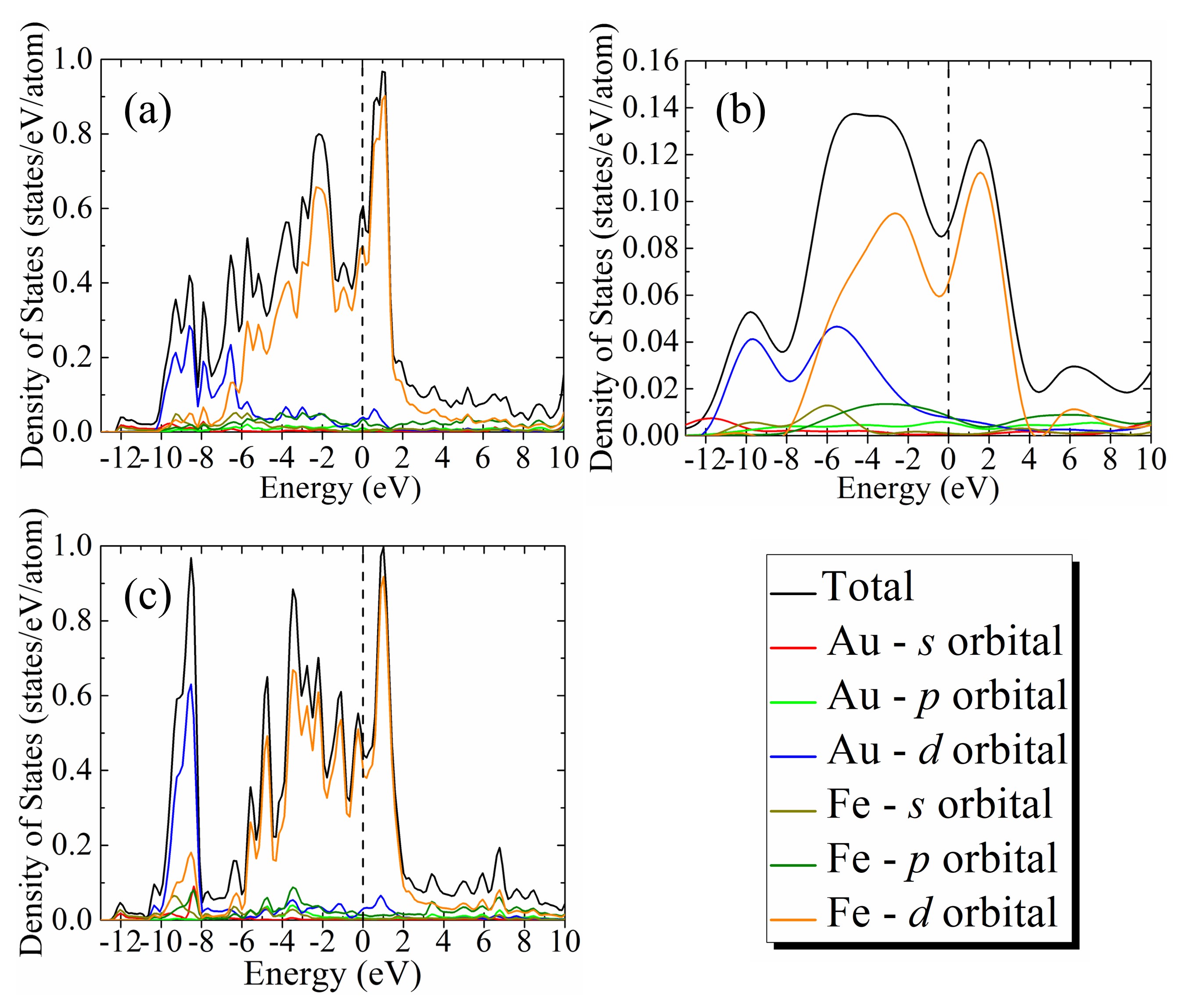}
	\caption{(color online) Calculated electronic density of states 
projected to orbitals for (a) Cmcm-AuFe$_{4}$ at 140 GPa, (b) 
Pmmn-AuFe$_{3}$ at 200 GPa, and (c) I4/m-AuFe$_{4}$ at 200 GPa. The 
black dashed line represents the Fermi energy level.}
	\label{fig:DOS}
\end{figure}

The calculated electronic band structures (Fig.\ S5) show that all the 
three Fe-Au systems reported in this work are metallic. The electronic 
densities of states (DOS) projected to orbitals (Fig.~\ref{fig:DOS}) 
reveal that the states in the vicinity of the Fermi energy level are 
primarily Fe 3d states, with some contribution from Au 5d states. This 
implies that the Fe 3d and Au 5d states are responsible for the 
metallicity of the phases reported here. 


The results of Bader charge analysis are shown in 
Tables~\ref{table:structure} and S2, as well as Fig.~S3. 
The observed atomic charges vary significantly between the three phases 
and, even within a given material, are highly dependent on the 
coordination environments. As expected from its greater 
electronegativity, Au acts as anion in the metastable Pmmn-AuFe$_{3}$ 
phase at 200 GPa, with each Au atom gaining 0.26e- and the Fe atoms at 
the edges of the Fe layers losing 0.19e-. The Fe atoms in the interior 
of the layers have greatly reduced partial charges of 0.03 and 0.04e-. 
A similar negative oxidation state in Au was reported for caesium auride 
(CsAu) \cite{biltz1938wertigkeit, sperling2008gold}. 

Conversely, for I4/m-AuFe$_{4}$ at 200 GPa, each Au atom lies at the 
center of the unit cell and is coordinated to 12 Fe atoms, in a geometry 
similar to what is observed in the ionic lattices of some actinide 
complexes such as [Th(NO$_{3}$)$_{6}$]$^{2-}$ \cite{corden1970crystal}.
Here, each Au atom is cationic, losing 0.40e-, while each Fe atom
gains 0.10e-, making them anionic. Fe is known to become highly 
electronegative at such high pressures and a similar charge-transfer 
mechanism has been reported for Fe and Ni in the Xe-Fe/Ni system 
\cite{stavrou2018synthesis}. 

Finally, for Cmcm-AuFe$_{4}$ at 140~GPa, the Au atoms each lose 0.34e-,
again making them cationic. However, the two distinct Fe environments 
in the crystal have very different atomic charges, similar to what was reported in $\alpha$-Mn \cite{magad2021high}. The Fe atoms at the 
first 8g sites directly bonded to the Au are also cationic and lose 
between 0.33e- and 0.37e-. Since they have comparable charges to the Au
atoms, we will refer to these Fe atoms as pseudo-Au atoms. The Fe 
atoms at the second 8g sites (in the interior of the Fe layer) gain 
between 0.51e- and 0.53e- each to become anions. Therefore, the 
Cmcm-AuFe$_{4}$ phase is stabilized through electron transfer from Au 
and pseudo-Au atoms to Fe atoms to give regions of alternating charge.

\subsection*{Dynamic, mechanical, and thermal properties of Fe-Au phases}

\begin{figure}
	\centering
	\includegraphics[width=1.0\linewidth]{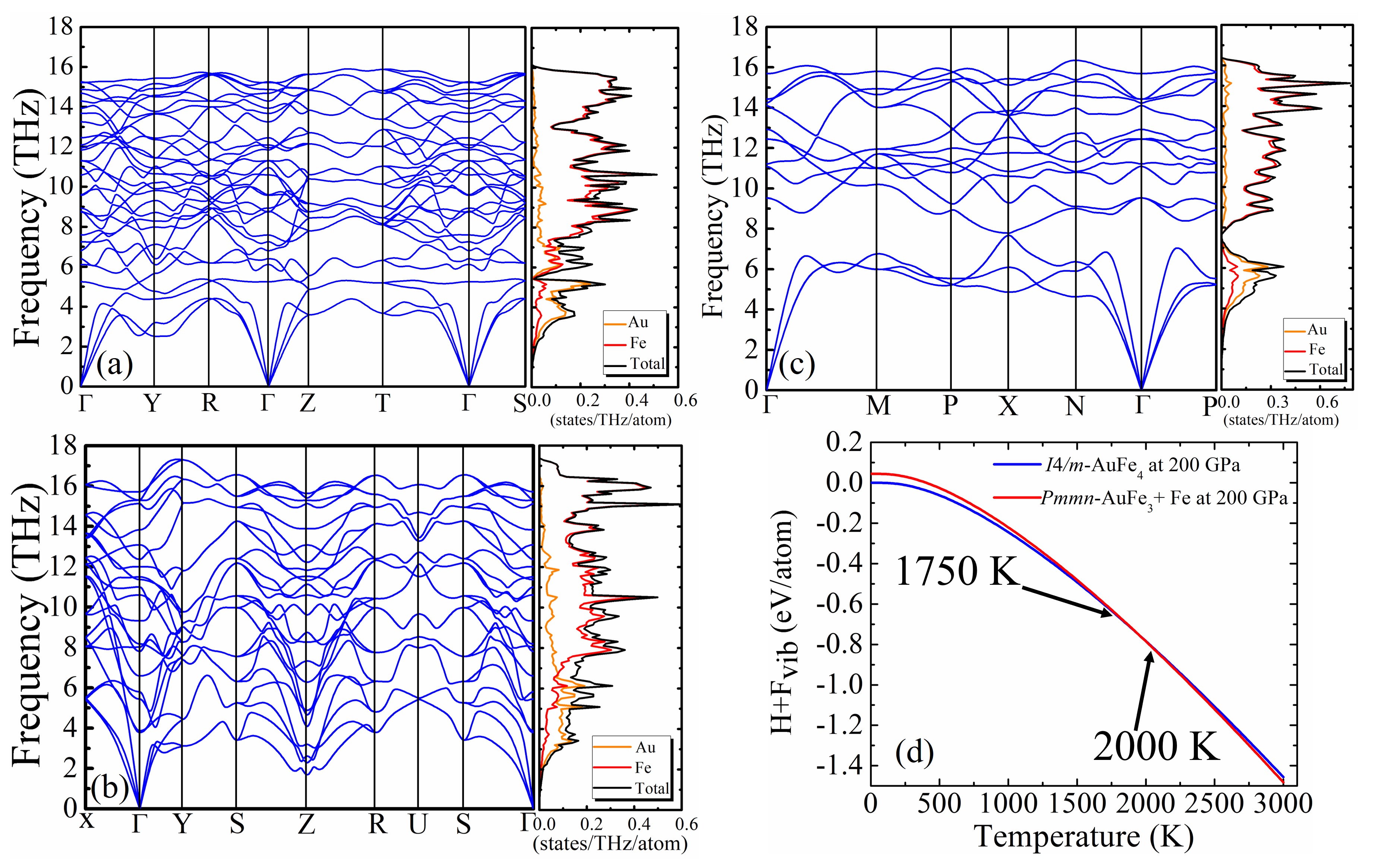}
	\caption{(color online) Phonon dispersion relations for (a) 
Cmcm-AuFe$_{4}$ at 140 GPa, (b) Pmmn-AuFe$_{3}$ at 200 GPa, and (c) 
I4/m-AuFe$_{4}$ at 200 GPa. Also shown is (d) the relative free 
energies of I4/m-AuFe$_{4}$ and Pmmn-AuFe$_{3}$ + hcp-Fe at 
200 GPa as a function of temperature. The I4/m-AuFe$_{4}$ structure at 
0~K was used as the zero of energy.}
	\label{fig:phonon}
\end{figure}

We established the dynamic stability of the various predicted phases of 
Fe-Au through the calculation of phonon dispersion relations 
(Fig.~\ref{fig:phonon} a-c). All three show no imaginary frequencies 
throughout the BZ, indicating dynamic stability within the harmonic 
approximation. We further investigated the response of the three 
predicted phases to external strain \cite{born1956theory,
hill1952elastic,hill1963elastic} which, in principle, is a measure of 
their elastic and mechanical stability (Table S1). The results indicate 
that all the predicted phases are elastically and mechanically stable 
and their Pugh's ratio \cite{mouhat2014necessary} also indicates that 
they are ductile.  

At 200 GPa, two structures are predicted to be synthesizable 
(Fig.~\ref{fig:EOS}b) with the I4/m-AuFe$_{4}$ phase being the 
thermodynamic ground state and the Pmmn-AuFe$_{3}$ phase being 
metastable. With an enthalpy difference of only $\sim$40 meV/atom 
between I4/m-AuFe$_{4}$ and the combination of the Pmmn-AuFe$_{3}$ 
phase and hcp-Fe, differences in their vibrational free energies at high 
temperature could compensate for the enthalpy difference and reverse the
stability ranking. Therefore, we calculated the Helmholtz free-energy 
evolution with temperature ($T$) at fixed volume ($V$) for both phases 
at 200 GPa within the harmonic approximation (Fig.~\ref{fig:phonon}d). 
The vibrational free-energy contribution is \cite{pavone1998alpha}
\begin{equation}\label{eqn:Fvib}
	F_{vib}(T)=\frac{1}{2}\sum_{\omega} g(\omega)\hbar\omega+k_{B}T\sum_{\omega}g(\omega)\ln\left[1-\exp\left(-\frac{\hbar\omega}{k_{B}T}\right)\right], 
\end{equation}
where g($\omega$) is the normalized density of states (phDOS) for the 
phonon branch $\omega$ and $k_{B}$ is Boltzmann's constant. 

The calculated temperature evolution of the relative free energies 
(Fig.~\ref{fig:phonon}d) reveals that I4/m-AuFe$_{4}$ is preferred at 
ambient temperature up to $\sim$1750~K. Between 1750 K and 2000 K, 
the two phases are effectively degenerate, while Pmmn-AuFe$_{3}$ + 
hcp-Fe becomes preferable at high temperatures above 2000~K. This 
preference can be motivated by the presence of more low-frequency 
phonons ($<$4 THz) for Pmmn-AuFe$_{3}$ (Fig.~\ref{fig:phonon}b) 
compared to I4/m-AuFe$_{4}$ (Fig.~\ref{fig:phonon}c), which will have 
a greater contribution to the vibrational entropy. The latent heat 
absorbed in the I4/m-AuFe$_{4}$ $\rightarrow$ Pmmn-AuFe$_{3}$ + hcp-Fe 
transition at 2000~K is estimated from the vibrational entropy to be 
0.025 eV/atom (Fig.~S9). The experimental implication of this 
observation is that the metastable Pmmn-AuFe$_{3}$ phase could 
potentially be prepared by laser-heating a compressed mixture of Au and 
Fe (with a molar ratio consistent with the stoichiometry) above 2000~K.

\subsection*{Thermal conductivity of Fe-Au}

The thermal conductivity of solids is governed by the phonon vibrations,
coupled with the scattering processes they encounter. During structural 
phase transitions driven by soft phonons, the phonon-assisted thermal 
conductivity is expected to be strongly modified, since the soft mode 
will experience a frequency and group velocity shift, which will in turn 
modify the allowed phonon scattering processes in the system. The 
modification of phonon modes also has significant implications for heat 
transport. 

\begin{figure}
\centering
\includegraphics[width=1.0\linewidth]{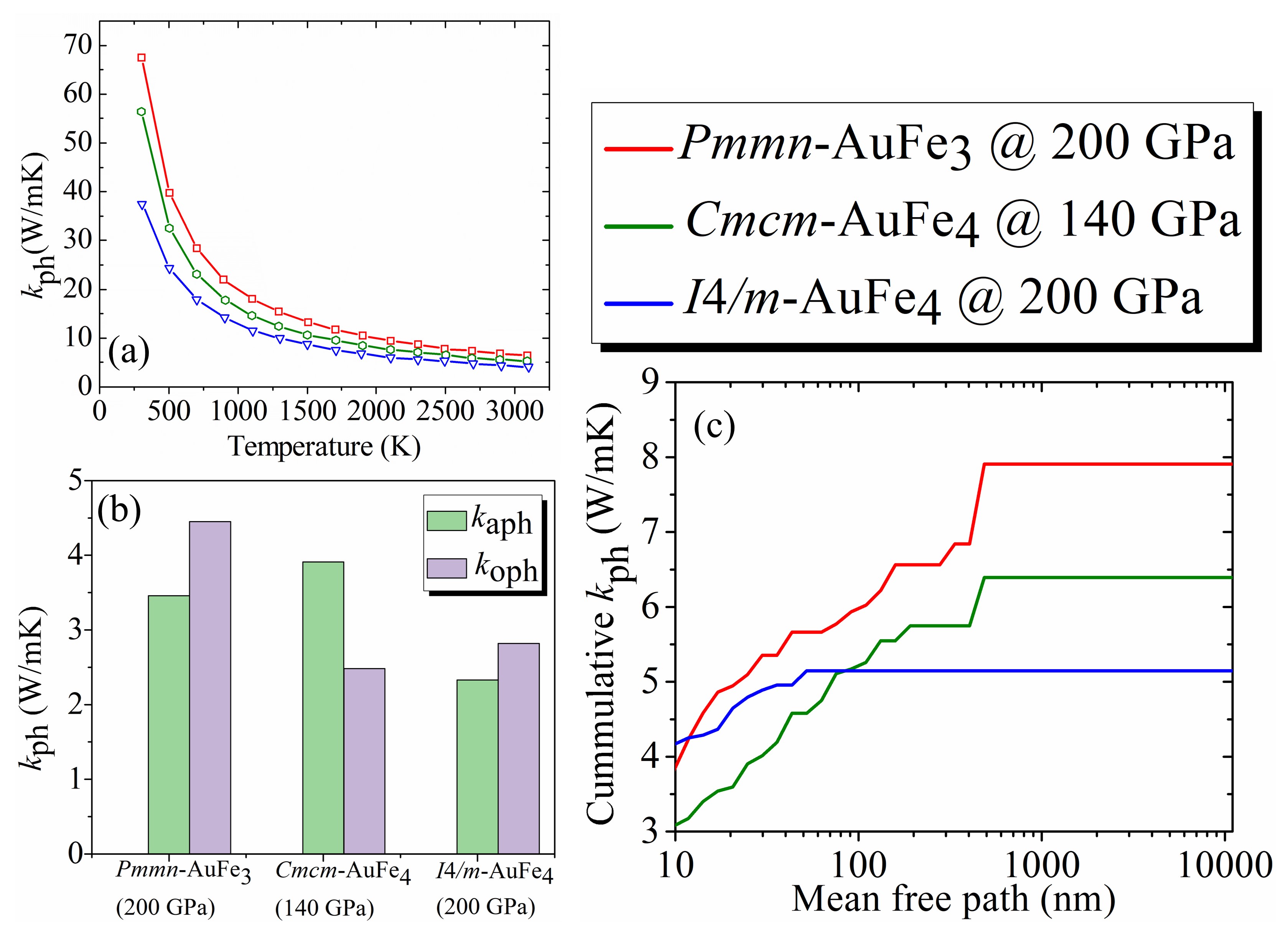}
\caption{(color online) (a) Calculated lattice thermal conductivity 
(k$_\text{ph}$) of the three Au-Fe phases as a function of temperature,
ranging from 300 to 3100 K. (b) Decomposition of the k$_\text{ph}$ of Au-Fe phases at 2500~K into contributions from the acoustic (k$_\text{aph}$) and 
optical (k$_\text{oph}$) modes. (c) Cumulative lattice thermal conductivity
of Cmcm-AuFe$_{4}$ (140 GPa), Pmmn-AuFe$_{3}$ (200 GPa), and 
I4/m-AuFe$_{4}$ (200 GPa) as a function of the phonon mean-free-path 
at 2500~K.}
\label{fig:Thermal_conduct}
\end{figure}

We computed the thermal conductivity of the three predicted Fe-Au 
intermetallic compounds by solving the Boltzmann transport equation (BTE) 
for a range of temperatures. The results are shown in 
Fig.~\ref{fig:Thermal_conduct}a and highlight the significant effect of 
phase change on the thermal conductivity of these compounds. While the 
relaxation time approximation (RTA) usually underestimates the thermal 
conductivity (since it treats both Normal and Umklapp scattering as 
resistive processes \cite{ziman2001electrons}), we found that the RTA 
produces results within ~0.5\% of the iterative solution for all three 
phases at room temperature. 


In the analysis of measured data, some \cite{guo2015anisotropic} have 
concluded that the acoustic modes dominate thermal transport and that 
the optical modes contribute negligibly due to low group velocities. 
However, we have found that this assumption cannot explain the 
comparatively large thermal conductivity ($k_\text{ph}$) values for the 
Fe-Au systems studied in this work. From Fig.~\ref{fig:phonon}a-c, 
there is no gap between the acoustic and optical modes in the phonon 
dispersion in either Cmcm-AuFe$_{4}$ or Pmmn-AuFe$_{3}$ at 200 GPa, 
while a small gap exists for I4/m-AuFe$_{4}$. Such gaps are often 
attributed to atomic mass differences and can suppress the interaction 
of acoustic phonons with optical phonons above the gap 
\cite{jain2014thermal}. To investigate further, we examined the 
contributions of each phonon mode from our BTE solution as shown in 
Fig.~\ref{fig:Thermal_conduct}c and Fig.~S4. At 300~K (Fig.~S4), 
optical phonon modes contribute 51\% of the total $k_\text{ph}$ of 
Pmmn-AuFe$_{3}$, while the contribution is 20\% and 43\% for 
Cmcm-AuFe$_{4}$ and I4/m-AuFe$_{4}$, respectively. At 2500~K 
(Fig.~\ref{fig:Thermal_conduct}c), their contribution increases 
for all three phases, to 56\%, 32\%, and 54\% for Pmmn-AuFe$_{3}$, Cmcm-AuFe$_{4}$, and I4/m-AuFe$_{4}$, 
respectively. This observation is consistent with the trend predicted 
from the group velocity ($v_{g}$) for acoustic modes, which would not 
support higher $k_\text{ph}$ without other contributions. The high
contributions from the optical modes are not surprising as similar 
contributions have been found in many complex materials 
\cite{guo2015thermal}. One reason for the strong optical phonon 
contribution may be due to the hybridization of low-energy optical 
modes with the acoustic modes. This causes a flattening of the acoustic 
dispersion for states away from the zone center (Fig. \ref{fig:phonon}) and, therefore, 
a reduction in $v_{g}$, leading to reduced thermal transport from the 
acoustic phonons. 

To determine the length scales for thermal transport, the cumulative 
$k_\text{ph}$ with respect to the phonon mean free path is shown in 
Fig.~\ref{fig:Thermal_conduct}b for the three phases of Au-Fe at 2500~K. 
This is informative for determining the length scales at which the 
phonons of each phase become relevant for heat conduction. Clearly, the 
I4/m-AuFe$_{4}$ phase is more sensitive to nano structuring compared to 
Pmmn-AuFe$_{3}$ and Cmcm-AuFe$_{4}$. This can be seen from the fact that the I4/m-AuFe$_{4}$ requires the least mean free path (< 100nm) to achieve its maximum, but the lowest lattice thermal conductivity and, as such, will 
transport heat more slowly. 

\section*{DISCUSSION}
\subsection*{Insight from elastic behavior of Fe-Au systems under pressure}

\begin{table*}
\centering
\caption{Calculated sound speed ($v_s$) and density at 200~GPa and 0~K 
for Pmmn-AuFe$_{3}$ and I4/m-AuFe$_{4}$ compared with simulation data 
\cite{li2018elastic} of inner Earth's core model at 360 GPa and the PREM 
\cite{dziewonski1981preliminary} data at 200~GPa.}
\begin{tabular}{lcccccc}
	\hline
	System & Pmmn-AuFe$_{3}$ & I4/m-AuFe$_{4}$ & hcp-Fe & Fe$_{62}$C$_{2}$ & Fe$_{60}$C$_{4}$ & PREM \\
	\midrule
	$v_{s}$ (km/s) & 4.28 & 4.82 & 6.80 & 6.38 & 6.0 & 3.6 \\
	Density (g/cm$^{3}$) & 17.72 & 16.88 & 14.25 & 14.10 & 13.82 & 13\\
	\bottomrule		
\end{tabular}
\label{table:elastic}
\end{table*}

The speed of elastic waves propagating through a material is a 
measurable property of condensed matter that is easily matched with 
seismic or experimental data when modeling the Earth's interior. Using 
Navier's relation \cite{anderson1963simplified,schreiber1975elastic}, we 
mapped the elastic properties space of the Fe-Au phases predicted in this 
study into longitudinal ($v_{l}$), transverse ($v_{t}$), and average 
($v_{m}$) elastic wave speeds. These speeds are:
\begin{equation}\label{eqn:transverse}
v_{t}=\sqrt{\frac{G}{\rho}},
\end{equation}
\begin{equation}\label{eqn:longitudinal}
v_{l}=\sqrt{\frac{3B+4G}{3\rho}} ,
\end{equation}
and
\begin{equation}\label{eqn:soundspeed}
v_{m}=\left[\frac{1}{3}\left(\frac{2}{v_{t}^{3}}+
\frac{1}{v_{l}^{3}}\right)\right]^{\frac{1}{3}} , 
\end{equation}
where $B$ and $G$ are the bulk and shear modulus, respectively, and 
$\rho$ is the density of the material. The average wave speed is 
reported as the speed of sound ($v_s$) in Table \ref{table:elastic}. 
The $v_s$ therefore captures the elastic anisotropy and, by extension, 
sound wave anisotropy in the system, should it exist.

Compounds in the Earth's interior each have different densities and 
sound speeds; thus, these values can be used in conjunction with 
seismic data to provide insight on its composition 
\cite{dziewonski1981preliminary, vovcadlo2007ab,kantor2007sound,
liu2014sound}. Seismological studies 
\cite{dewaele2006quasihydrostatic} suggest that the Earth's core must 
include some light siderophile elements to explain its lower density 
than that of pure Fe \cite{allegre1995chemical, adeleke2020formation,
adeleke2020two,poirier1994light}. In addition to having a higher 
density, hcp-Fe has a much higher velocity of sound than predicted 
from models of the core, as shown in Table~\ref{table:elastic}. 
While introducing light elements, as in Fe$_{62}$C$_2$ and 
Fe$_{60}$C$_4$ \cite{li2018elastic}, improves agreement with the PREM 
density, it does so at the expense of worsening agreement in the sound 
speed.

The extracted density and sound speeds for the Pmmn-AuFe$_{4}$ and 
I4/m-AuFe$_{3}$ phases are also reported in Table~\ref{table:elastic}.
The Cmcm-AuFe$_{4}$ phase was not included 
because the pressure at which we are making this comparison is 
outside the stability range of this phase (Fig.~\ref{fig:EOS}c). 
Compared with hcp-Fe or other Fe-rich model structures (Fe$_{62}$C$_2$ 
or Fe$_{60}$C$_4$) \cite{li2018elastic}, Pmmn-AuFe$_{3}$ and 
I4/m-AuFe$_{4}$ have average sound speeds that are very much closer 
to that from the PREM model \cite{dziewonski1981preliminary}. Particularly, the metastable Pmmn-AuFe$_{3}$ phase has $v_{s}$ that is closest to the PREM's, further supporting the stabilization of the Pmmn-AuFe$_{3}$ over the I4/m-AuFe$_{4}$ at the Earth's core's (> 2000~K) temperature. However, 
the densities of Pmmn-AuFe$_{4}$ and I4/m-AuFe$_{3}$ are significantly
higher than both hcp-Fe and the PREM model data. The results suggest that combining Fe and Au with Ni or other light elements 
to form ternary systems could yield calculated densities and sound 
speeds closer to the PREM data \cite{he2022superionic}. Overall, the addition of Au to the base Fe leads to a decrease in seismic velocities towards the geophysical observations at the CMB and the outer core. We opine that, if Au is present, the Fe 
content that will be required to match precisely the Earth's core profile 
would be less than what is required in the light-element-Fe alloys 
\cite{li2018elastic}.

\subsection*{Implications for thermal conductivity at the CMB}

The thermal conductivities of the Earth's lower mantle and core greatly 
impact convection dynamics. They also determine the ease of heat 
transport from the core to the mantle, thereby maintaining the heat 
energy budget for the Earth's geodynamo. As a result, there is a large 
body of work investigating thermal properties of iron and its alloys 
within a pressure-temperature regime relevant to the silicate mantle and 
core \cite{yong2019iron,gomi2013high,de2012electrical}. Meanwhile, 
direct measurement of thermal conductivity under such extreme conditions 
still poses formidable challenges, so estimation of thermal 
conductivities from first-principles electronic structure calculations 
must suffice. Thus, identifying alloys and compounds of Fe with 
potentially high thermal conductivity is interesting as they could 
facilitate thermal convection at both the core-mantle boundary (CMB) 
and the inner-core boundary (ICB). 

The coupling between mobile electrons and lattice vibrations dominates 
the heat transfer process of metals. The Wiedemann-Franz law 
\cite{anderson1998gruneisen} describes the relationship between 
electronic thermal conductivity and electrical resistivity as:
\begin{equation}\label{eqn:k-total}
	k = \frac{LT}{\varrho}, 
\end{equation}
where $L=2.44 \times 10^{-8} W\Omega / K^{2}$ is the Lorentz number, 
$T$ is the absolute temperature, and $\varrho$ is the electrical 
resistivity. Thus, calculation of the electronic thermal conductivity 
of Fe-Au compounds is dependent on the knowledge of their electrical 
resistivity, which is scarce in the literature, especially for newly 
synthesized or predicted materials, and its determination falls beyond 
the scope of this paper. However, the Wiedemann-Franz law is a lower 
bound for the total thermal conductivity in a metal \cite{gomi2013high}.
While the addition of the electronic contribution would help to 
quantitatively estimate the total thermal conductivity of the Fe-Au 
systems, calculation of the total phonon thermal conductivity 
(including both acoustic and optical processes) gives a reasonable 
description of the heat transport mechanism \cite{gomi2013high}. 

The total phonon thermal conductivity ($k_\text{ph}$) in the predicted 
Fe-Au compounds could be as high as 68~W/mK in Pmmn-AuFe$_{3}$ at 
200~GPa and room temperature, and as low as 5.2~W/mK in I4/m-AuFe$_{4}$ 
at 200~GPa and 2500~K (Fig.~\ref{fig:Thermal_conduct}a). Notably, the 
thermal conductivity of the CMB, corresponding to 136~GPa and 3750~K, is reported to be 90~W/mK \cite{gomi2013high}. The pressure regime where 
Cmcm-AuFe$_{4}$ is stable is therefore most representative of the CMB. 
Cmcm-AuFe$_{4}$ has a $k_\text{ph}$ of 6 W/mK at 140~GPa and 
3100~K. The $k_\text{ph}$ of the Cmcm-AuFe$_{4}$ corresponds to 6.7\% of the total thermal conductivity of the CMB. Given that the reported thermal conductivity in Ref.~\cite{gomi2013high} already captured the electronic and 
vibrational contributions to the thermal conductivity (as well as 
those from any impurities), it then implies that the phonon process is 
significant in the Fe-Au compounds during heat transport at the CMB.
Excess heat that is not transported by the mantle- or CMB-bound Fe-Au compounds 
is, instead, transported through compositional convection into the 
inner core \cite{gomi2013high}. Unreacted Au, if available, may be 
transported through this process into the solid Earth's inner core to 
form other stable Fe-Au compounds, such as Pmmn-AuFe$_{3}$ and 
I4/m-AuFe$_{4}$. 


\section*{CONCLUSION}

In this work, we investigated the possible formation of stable Fe-Au 
compounds under pressure using density-functional methods. We found 
that, at high pressure, Au forms several stable compounds with Fe, 
which has strong implication for understanding the form that Au may 
take in the Earth's mantle, CMB, and beyond. The Fe-Au intermetallic 
compounds are predicted to be stabilized by high pressure and electron 
transfer. Au exists as an anion in AuFe$_{3}$ and as cation in two 
AuFe$_{4}$ polymorphs, providing further evidence to support the 
striking chemistry of Au being able to adopt variable oxidation states
at high pressure. At 140~GPa, Au attains a magnetic moment of 
0.52$\mu_{B}$ in Cmcm-AuFe$_{4}$, making it ferromagnetic in nature. The speed of sound calculated for the various Fe-Au phases shows lower deviation from the PREM data compared to the other Fe-light element models compared. 
Furthermore, the computed thermal conductivity data for the various
Fe-Au compounds shows that Fe-Au compounds have significant lattice thermal conductivity. This suggests that, beyond electronic phenomenon, the lattice vibrations also contribute to heat transport within the Earth's core and at the CMB. The results from this study show that some part of the depleted Au could be mixed with Fe in the CMB and the outer region of the core. As such, our Fe-Au model could serve to explain the seismological structure of the upper part of the Earth's interior, up to the core-mantle boundary, and some part of the outer core.  Our work has strong implications for geoscience in the area of the Earth's magnetic field and could be extended to study how Au is stored in the Earth's solid, ferric inner core at pressures above 300~GPa.

\section*{MATERIALS AND METHODS}
The crystal structure search was carried out using the particle 
swarm-intelligence optimization (PSO) algorithm using the Calypso 
program \cite{wang2010crystal,wang2012calypso}. The structure search 
was performed for pressures of 0, 50, 100, 140, and 200 GPa, with 
simulation cells containing 1-4 formula units of Fe$_{x}$Au$_{y}$. 
The stability of various phases with respect to decomposition was 
assessed by constructing their convex hull. Electronic structure 
calculations for structural optimization and property evaluations
were performed with density-functional theory (DFT) and dynamic 
stability calculations with density-functional perturbation theory 
(DFPT) \cite{baroni2001phonons} as implemented in the Vienna Ab initio 
Simulation Package (VASP) \cite{kresse1993ab} code. These calculations
used the projector-augmented wave (PAW) \cite{kresse1999ultrasoft} 
approach, in which the valence states of Au and Fe were treated as 
5d$^{10}$6s$^{1}$ and 3p$^{6}$3d$^{7}$4s$^{1}$, respectively. The
Perdew-Burke-Ernzerhof (PBE) \cite{perdew1996generalized} generalized 
gradient approximation (GGA) functional was selected and the planewave
energy cut-off was set to 450 eV. The GGA+$U$ framework was used to 
apply an on-site Coulomb interaction to improve treatment of the 3d 
electrons of Fe. The Hubbard $U$ parameter was set to 8.6 eV 
\cite{cococcioni2005linear, majumdar2017superconductivity} and such 
treatment of Fe has previously shown good agreement with experimental 
results \cite{mao1990static} over the pressure range of interest. To 
ensure that forces on all atoms were converged to within 1 meV/\r{A}, 
the Monkhorst-Pack scheme was used to sample the Brillouin zone, with a \textbf{k} spacing of 2$\pi$ $\times$ 0.02 \r{A}$^{-1}$.
Dynamical calculations for the thermal properties were carried out 
using the Quantum Espresso (QE) code \cite{giannozzi2009quantum}, 
again with the PBE+$U$ exchange-correlation functional. Well converged total energies 
were obtained using a planewave kinetic-energy cut-off of 150 Ry and 
Monkhorst-Pack \textbf{k}-point meshes of $9 \times 9\times 3$ and 
$8 \times 8\times 8$ for AuFe$_3$ and AuFe$_4$, respectively. 

To predict the thermal conductivity, the ShengBTE code 
\cite{li2014shengbte} was used to iteratively solve the Boltzmann 
Transport Equation (BTE):
\begin{equation}\label{eqn:f-lamb}
F_{\lambda}=\tau_{\lambda}^{0}\left(V_{\lambda}+\Delta_{\lambda}\right)  .
\end{equation}
Here, $F_{\lambda}$ is the generalized mean free path, 
$\tau_{\lambda}^{0}$ is the relaxation time of mode $\lambda$ in the 
relaxation time approximation (RTA), and $\Delta_{\lambda}$ gives the 
deviation of the solution from the RTA. The phonon assisted thermal 
conductivity ($k_\text{ph}$) tensor can then be obtained as
\begin{equation}\label{eqn:k-ph}
	k_\text{ph}^{\alpha\beta}=\frac{1}{k_{B}T^{2}\Omega N}\sum_{\lambda}n^{0}(n^{0}+)(\hbar\omega_{\lambda})^{2}v_{\alpha,\lambda}F_{\beta,\lambda}  ,
\end{equation}
where $\alpha$ and $\beta$ are the three coordinate directions 
($x$, $y$, and $z$). $k_{B}$, $T$, $\Omega$, and $N$ are Boltzmann's
constant, the temperature, the unit-cell volume, and the number of 
\textbf{q}-points in the integral over the BZ, respectively. The sum 
runs over all the phonon modes $\lambda$, $\hbar$ is the reduced Planck 
constant, and $\omega_{\lambda}$ is the phonon frequency. For the first 
iteration step, $\Delta\lambda$ is set to zero, which is equivalent to 
starting the iterative procedure from the RTA solution. The solution is 
converged when the relative change in all the components of the thermal 
conductivity tensor becomes smaller than $10^{-5}$~W/mK. 
To solve the BTE, we used a $4 \times 4 \times 1$ supercell to calculate 
the third-order interatomic force constants (IFCs) for the Pmmn and Cmcm 
phases, and a $3 \times 3 \times 3$ supercell for the I4/m phase. The 
force cut-off distance was set such that the interaction range includes 
the five nearest neighbors for AuFe$_{3}$ and the three nearest neighbors 
for AuFe$_{4}$. Meshes of $7 \times 7 \times 5$ and $5 \times 5 \times 5$ 
\textbf{q}-points were used to calculate the second-order IFCs needed to 
compute the $k_\text{ph}$ of AuFe$_{3}$ and AuFe$_{4}$, respectively.




\subsection*{Data Availability}
All study data are included in the article and/or SI Appendix.

\subsection*{Supporting Information Appendix (SI)}
The supporting information contains additional computational results, including plots of the zero-point energy (ZPE)-corrected convex hull, equation of states, 2D-projected crystal structures with Bader charge distributions, thermal conductivity at 300 K, electronic band structures, projected electronic density of states, vibrational entropies evolution with temperature, elastic modulus data and atomic charge distributions. 


\acknow{A.A.A. and E.R.J acknowledge the support by the Natural Sciences and Engineering Research Council of Canada (NSERC). Computing resource is provided by Lawrence Livermore National Laboratory, Westgrid, and Compute Canada.}

\showacknow{} 

\subsection*{References}
\bibliographystyle{plain}
\bibliography{pnas-new}

\begin{thebibliography}{69}
\providecommand{\natexlab}[1]{#1}
\providecommand{\url}[1]{\texttt{#1}}
\expandafter\ifx\csname urlstyle\endcsname\relax
  \providecommand{\doi}[1]{doi: #1}\else
  \providecommand{\doi}{doi: \begingroup \urlstyle{rm}\Url}\fi

\bibitem[Wood et~al.(2006)Wood, Walter, and Wade]{wood2006accretion}
Bernard~J Wood, Michael~J Walter, and Jonathan Wade.
\newblock Accretion of the earth and segregation of its core.
\newblock \emph{Nature}, 441\penalty0 (7095):\penalty0 825--833, 2006.

\bibitem[Stevenson(1981)]{stevenson1981models}
DJ~Stevenson.
\newblock Models of the earth's core.
\newblock \emph{Science}, 214\penalty0 (4521):\penalty0 611--619, 1981.

\bibitem[Jones(1968)]{jones1968gold}
Robert~Sprague Jones.
\newblock \emph{Gold in Meteorites and in the Earth's Crust}, volume 603.
\newblock US Government Printing Office, 1968.

\bibitem[McDonough and Sun(1995)]{mcdonough1995composition}
William~F McDonough and S-S Sun.
\newblock The composition of the earth.
\newblock \emph{Chemical Geology}, 120\penalty0 (3-4):\penalty0 223--253, 1995.

\bibitem[Ahrens(1995)]{ahrens1995global}
Thomas~J Ahrens.
\newblock \emph{Global earth physics: a handbook of physical constants},
  volume~1.
\newblock American Geophysical Union, 1995.

\bibitem[Helffrich and Wood(2001)]{helffrich2001earth}
George~R Helffrich and Bernard~J Wood.
\newblock The earth's mantle.
\newblock \emph{Nature}, 412\penalty0 (6846):\penalty0 501--507, 2001.

\bibitem[McKay and Peters(2017)]{mckay2017midas}
Douglas~R McKay and Daniel~A Peters.
\newblock The midas touch: Gold and its role in the global economy.
\newblock \emph{Plastic Surgery}, 25\penalty0 (1):\penalty0 61--63, 2017.

\bibitem[Briggs et~al.(2019)Briggs, Coppari, Gorman, Smith, Tracy, Coleman,
  Fernandez-Pa{\~n}ella, Millot, Eggert, and
  Fratanduono]{briggs2019measurement}
R~Briggs, F~Coppari, MG~Gorman, RF~Smith, SJ~Tracy, AL~Coleman,
  A~Fernandez-Pa{\~n}ella, M~Millot, JH~Eggert, and DE~Fratanduono.
\newblock Measurement of body-centered cubic gold and melting under shock
  compression.
\newblock \emph{Physical Review Letters}, 123\penalty0 (4):\penalty0 045701,
  2019.

\bibitem[Dubrovinsky et~al.(2007)Dubrovinsky, Dubrovinskaia, Crichton,
  Mikhaylushkin, Simak, Abrikosov, de~Almeida, Ahuja, Luo, and
  Johansson]{dubrovinsky2007noblest}
Leonid Dubrovinsky, Natalia Dubrovinskaia, Wilson~A Crichton, Arkady~S
  Mikhaylushkin, Sergey~I Simak, Igor~A Abrikosov, J~Souza de~Almeida, Rajeev
  Ahuja, Wei Luo, and B{\"o}rje Johansson.
\newblock Noblest of all metals is structurally unstable at high pressure.
\newblock \emph{Physical Review Letters}, 98\penalty0 (4):\penalty0 045503,
  2007.

\bibitem[Ishikawa et~al.(2013)Ishikawa, Kato, Nomura, Suzuki, Nagara, and
  Shimizu]{ishikawa2013pressure}
Takahiro Ishikawa, Kuniko Kato, Masaya Nomura, Naoshi Suzuki, Hitose Nagara,
  and Katsuya Shimizu.
\newblock Pressure-induced stacking sequence variations in gold from first
  principles.
\newblock \emph{Physical Review B}, 88\penalty0 (21):\penalty0 214110, 2013.

\bibitem[Ahuja et~al.(2001)Ahuja, Rekhi, and Johansson]{ahuja2001theoretical}
Rajeev Ahuja, Sandeep Rekhi, and B{\"o}rje Johansson.
\newblock Theoretical prediction of a phase transition in gold.
\newblock \emph{Physical Review B}, 63\penalty0 (21):\penalty0 212101, 2001.

\bibitem[S{\"o}derlind(2002)]{soderlind2002comment}
Per S{\"o}derlind.
\newblock Comment on ``theoretical prediction of phase transition in gold''.
\newblock \emph{Physical Review B}, 66\penalty0 (17):\penalty0 176201, 2002.

\bibitem[Boettger(2003)]{boettger2003theoretical}
JC~Boettger.
\newblock Theoretical extension of the gold pressure calibration standard
  beyond 3 mbars.
\newblock \emph{Physical Review B}, 67\penalty0 (17):\penalty0 174107, 2003.

\bibitem[Wasson(1985)]{wasson1985meteorites}
John~T Wasson.
\newblock Meteorites: their record of early solar-system history.
\newblock \emph{New York: Freeman}, 1985.

\bibitem[Dziewonski and Anderson(1981)]{dziewonski1981preliminary}
Adam~M Dziewonski and Don~L Anderson.
\newblock Preliminary reference earth model.
\newblock \emph{Physics of the Earth and Planetary Interiors}, 25\penalty0
  (4):\penalty0 297--356, 1981.

\bibitem[Tkal{\v{c}}i{\'c} and Ph\d{a}m(2018)]{tkalvcic2018shear}
Hrvoje Tkal{\v{c}}i{\'c} and Thanh-Son Ph\d{a}m.
\newblock Shear properties of earth's inner core constrained by a detection of
  j waves in global correlation wavefield.
\newblock \emph{Science}, 362\penalty0 (6412):\penalty0 329--332, 2018.

\bibitem[Norris and Wood(2017)]{norris2017earth}
C~Ashley Norris and Bernard~J Wood.
\newblock Earth's volatile contents established by melting and vaporization.
\newblock \emph{Nature}, 549\penalty0 (7673):\penalty0 507--510, 2017.

\bibitem[Dewaele et~al.(2006)Dewaele, Loubeyre, Occelli, Mezouar, Dorogokupets,
  and Torrent]{dewaele2006quasihydrostatic}
Agnes Dewaele, Paul Loubeyre, Florent Occelli, Mohamed Mezouar, Peter~I
  Dorogokupets, and Marc Torrent.
\newblock Quasihydrostatic equation of state of iron above 2 mbar.
\newblock \emph{Physical Review Letters}, 97\penalty0 (21):\penalty0 215504,
  2006.

\bibitem[Allegre et~al.(1995)Allegre, Poirier, Humler, and
  Hofmann]{allegre1995chemical}
Claude~J Allegre, Jean-Paul Poirier, Eric Humler, and Albrecht~W Hofmann.
\newblock The chemical composition of the earth.
\newblock \emph{Earth and Planetary Science Letters}, 134\penalty0
  (3-4):\penalty0 515--526, 1995.

\bibitem[Adeleke and Yao(2020)]{adeleke2020formation}
Adebayo~A Adeleke and Yansun Yao.
\newblock Formation of stable compounds of potassium and iron under pressure.
\newblock \emph{Journal of Physical Chemistry A}, 124\penalty0 (23):\penalty0
  4752--4763, 2020.

\bibitem[Adeleke et~al.(2020)Adeleke, Stavrou, Adeniyi, Wan, Gou, and
  Yao]{adeleke2020two}
Adebayo~A Adeleke, Elissaios Stavrou, Adebayo~O Adeniyi, Biao Wan, Huiyang Gou,
  and Yansun Yao.
\newblock Two good metals make a semiconductor: A potassium-nickel compound
  under pressure.
\newblock \emph{Physical Review B}, 102\penalty0 (13):\penalty0 134120, 2020.

\bibitem[Poirier(1994)]{poirier1994light}
Jean-Paul Poirier.
\newblock Light elements in the earth's outer core: a critical review.
\newblock \emph{Physics of the Earth and Planetary Interiors}, 85\penalty0
  (3-4):\penalty0 319--337, 1994.

\bibitem[Stavrou et~al.(2018)Stavrou, Yao, Goncharov, Lobanov, Zaug, Liu,
  Greenberg, and Prakapenka]{stavrou2018synthesis}
Elissaios Stavrou, Yansun Yao, Alexander~F Goncharov, Sergey~S Lobanov,
  Joseph~M Zaug, Hanyu Liu, Eran Greenberg, and Vitali~B Prakapenka.
\newblock Synthesis of xenon and iron-nickel intermetallic compounds at earth's
  core thermodynamic conditions.
\newblock \emph{Physical Review Letters}, 120\penalty0 (9):\penalty0 096001,
  2018.

\bibitem[Adeleke et~al.(2019)Adeleke, Kunz, Greenberg, Prakapenka, Yao, and
  Stavrou]{adeleke2019high}
Adebayo~A Adeleke, Martin Kunz, Eran Greenberg, Vitali~B Prakapenka, Yansun
  Yao, and Elissaios Stavrou.
\newblock A high-pressure compound of argon and nickel: Noble gas in the
  earth's core?
\newblock \emph{ACS Earth and Space Chemistry}, 3\penalty0 (11):\penalty0
  2517--2524, 2019.

\bibitem[Nielsen et~al.(1993)Nielsen, Besenbacher, Stensgaard, Laegsgaard,
  Engdahl, Stoltze, Jacobsen, and N{\o}rskov]{nielsen1993initial}
L~Pleth Nielsen, Flemming Besenbacher, I~Stensgaard, E~Laegsgaard, C~Engdahl,
  Per Stoltze, Karsten~Wedel Jacobsen, and Jens~Kehlet N{\o}rskov.
\newblock Initial growth of au on ni (110): Surface alloying of immiscible
  metals.
\newblock \emph{Physical Review Letters}, 71\penalty0 (5):\penalty0 754, 1993.

\bibitem[Mehendale et~al.(2010)Mehendale, Girard, Repain, Chacon, Lagoute,
  Rousset, Marathe, and Narasimhan]{mehendale2010ordered}
S~Mehendale, Y~Girard, V~Repain, C~Chacon, J~Lagoute, S~Rousset, Madhura
  Marathe, and Shobhana Narasimhan.
\newblock Ordered surface alloy of bulk-immiscible components stabilized by
  magnetism.
\newblock \emph{Physical Review Letters}, 105\penalty0 (5):\penalty0 056101,
  2010.

\bibitem[Dong et~al.(2015)Dong, Oganov, Qian, Zhou, Zhu, and
  Wang]{dong2015chemical}
Xiao Dong, Artem~R Oganov, Guangrui Qian, Xiang-Feng Zhou, Qiang Zhu, and
  Hui-Tian Wang.
\newblock How do chemical properties of the atoms change under pressure.
\newblock \emph{arXiv preprint arXiv:1503.00230}, 2015.

\bibitem[Tang et~al.(2009)Tang, Sanville, and Henkelman]{tang2009grid}
W~Tang, E~Sanville, and G~Henkelman.
\newblock A grid-based bader analysis algorithm without lattice bias.
\newblock \emph{Journal of Physics: Condensed Matter}, 21\penalty0
  (8):\penalty0 084204, 2009.

\bibitem[Jansen(2008)]{jansen2008chemistry}
Martin Jansen.
\newblock The chemistry of gold as an anion.
\newblock \emph{Chemical Society Reviews}, 37\penalty0 (9):\penalty0
  1826--1835, 2008.

\bibitem[Shih(1931)]{shih1931magnetic}
Ju~Wei Shih.
\newblock Magnetic properties of gold-iron alloys.
\newblock \emph{Physical Review}, 38\penalty0 (11):\penalty0 2051, 1931.

\bibitem[Kaufmann et~al.(1945)Kaufmann, Pan, and
  Clark]{kaufmann1945magnetization}
AR~Kaufmann, ST~Pan, and JR~Clark.
\newblock Magnetization of gold-iron and gold-nickel solid solutions.
\newblock \emph{Reviews of Modern Physics}, 17\penalty0 (1):\penalty0 87, 1945.

\bibitem[Pan et~al.(1942)Pan, Kaufmann, and Bitter]{pan1942ferromagnetic}
ST~Pan, AR~Kaufmann, and F~Bitter.
\newblock Ferromagnetic gold-iron alloys.
\newblock \emph{Journal of Chemical Physics}, 10\penalty0 (6):\penalty0
  318--321, 1942.

\bibitem[Stamatelatos et~al.(2019)Stamatelatos, Poulopoulos, Goschew,
  Fumagalli, Sarigiannidou, Rapenne, Opagiste, Grammatikopoulos, Wilhelm, and
  Rogalev]{stamatelatos2019paramagnetic}
A~Stamatelatos, P~Poulopoulos, A~Goschew, P~Fumagalli, E~Sarigiannidou,
  L~Rapenne, Christine Opagiste, S~Grammatikopoulos, F~Wilhelm, and A~Rogalev.
\newblock Paramagnetic gold in a highly disordered {Au-Ni-O} alloy.
\newblock \emph{Scientific Reports}, 9\penalty0 (1):\penalty0 1--8, 2019.

\bibitem[Zheng-Johansson et~al.(1998)Zheng-Johansson, Eriksson, Johansson,
  Fast, and Ahuja]{zheng1998comment}
JX~Zheng-Johansson, O~Eriksson, B~Johansson, L~Fast, and R~Ahuja.
\newblock Comment on ``stability and the equation of state of
  $\alpha$-manganese under ultrahigh pressure''.
\newblock \emph{Physical Review B}, 57\penalty0 (17):\penalty0 10989, 1998.

\bibitem[Biltz et~al.(1938)Biltz, Weibke, Hans-Joachim~Ehrhorn, Wedemeyer, and
  Weibke]{biltz1938wertigkeit}
Wilhelm Biltz, Friedrich Weibke, Versuchen Hans-Joachim~Ehrhorn, Roman
  Wedemeyer, and Friedrich Weibke.
\newblock {\"U}ber wertigkeit und chemische kompression von metallen in
  verbindung mit gold.
\newblock \emph{Zeitschrift f{\"u}r anorganische und allgemeine Chemie},
  236\penalty0 (1):\penalty0 12--23, 1938.

\bibitem[Sperling et~al.(2008)Sperling, Gil, Zhang, Zanella, and
  Parak]{sperling2008gold}
RA~Sperling, PR~Gil, F~Zhang, M~Zanella, and WJ~Parak.
\newblock Gold: Chemistry, materials and catalysis issue.
\newblock \emph{Chemical Society Review}, 37\penalty0 (9):\penalty0 1909--1930,
  2008.

\bibitem[Corden et~al.(1970)Corden, Cunningham, and
  Eisenberg]{corden1970crystal}
Brian~J Corden, James~A Cunningham, and Richard Eisenberg.
\newblock Crystal and molecular structure of tris (tetra-n-butylammonium)
  octacyanomolybdate (v).
\newblock \emph{Inorganic Chemistry}, 9\penalty0 (2):\penalty0 356--362, 1970.

\bibitem[Magad-Weiss et~al.(2021)Magad-Weiss, Adeleke, Greenberg, Prakapenka,
  Yao, and Stavrou]{magad2021high}
Logan~K Magad-Weiss, Adebayo~A Adeleke, Eran Greenberg, Vitali~B Prakapenka,
  Yansun Yao, and Elissaios Stavrou.
\newblock High-pressure structural study of $\alpha$-mn: Experiments and
  calculations.
\newblock \emph{Physical Review B}, 103\penalty0 (1):\penalty0 014101, 2021.

\bibitem[Born and Huang(1956)]{born1956theory}
M~Born and K~Huang.
\newblock Theory of crystal lattices, clarendon, 1956.

\bibitem[Hill(1952)]{hill1952elastic}
Richard Hill.
\newblock The elastic behaviour of a crystalline aggregate.
\newblock \emph{Proceedings of the Physical Society. Section A}, 65\penalty0
  (5):\penalty0 349, 1952.

\bibitem[Hill(1963)]{hill1963elastic}
Rodney Hill.
\newblock Elastic properties of reinforced solids: some theoretical principles.
\newblock \emph{Journal of the Mechanics and Physics of Solids}, 11\penalty0
  (5):\penalty0 357--372, 1963.

\bibitem[Mouhat and Coudert(2014)]{mouhat2014necessary}
F{\'e}lix Mouhat and Fran{\c{c}}ois-Xavier Coudert.
\newblock Necessary and sufficient elastic stability conditions in various
  crystal systems.
\newblock \emph{Physical Review B}, 90\penalty0 (22):\penalty0 224104, 2014.

\bibitem[Pavone et~al.(1998)Pavone, Baroni, and de~Gironcoli]{pavone1998alpha}
Pasquale Pavone, Stefano Baroni, and Stefano de~Gironcoli.
\newblock $\alpha \leftrightarrow \beta$ phase transition in tin: A theoretical
  study based on density-functional perturbation theory.
\newblock \emph{Physical Review B}, 57\penalty0 (17):\penalty0 10421, 1998.

\bibitem[Ziman(2001)]{ziman2001electrons}
John~M Ziman.
\newblock \emph{Electrons and phonons: the theory of transport phenomena in
  solids}.
\newblock Oxford university press, 2001.

\bibitem[Guo et~al.(2015{\natexlab{a}})Guo, Verma, Wu, Sun, Hickman, Masui,
  Kuramata, Higashiwaki, Jena, and Luo]{guo2015anisotropic}
Zhi Guo, Amit Verma, Xufei Wu, Fangyuan Sun, Austin Hickman, Takekazu Masui,
  Akito Kuramata, Masataka Higashiwaki, Debdeep Jena, and Tengfei Luo.
\newblock Anisotropic thermal conductivity in single crystal $\beta$-gallium
  oxide.
\newblock \emph{Applied Physics Letters}, 106\penalty0 (11):\penalty0 111909,
  2015{\natexlab{a}}.

\bibitem[Jain and McGaughey(2014)]{jain2014thermal}
Ankit Jain and Alan~JH McGaughey.
\newblock Thermal conductivity of compound semiconductors: Interplay of mass
  density and acoustic-optical phonon frequency gap.
\newblock \emph{Journal of Applied Physics}, 116\penalty0 (7):\penalty0 073503,
  2014.

\bibitem[Guo et~al.(2015{\natexlab{b}})Guo, Wang, and Huang]{guo2015thermal}
Ruiqiang Guo, Xinjiang Wang, and Baoling Huang.
\newblock Thermal conductivity of skutterudite {CoSb}$_3$ from first
  principles: substitution and nanoengineering effects.
\newblock \emph{Scientific Reports}, 5\penalty0 (1):\penalty0 1--9,
  2015{\natexlab{b}}.

\bibitem[Li et~al.(2018)Li, Vo{\v{c}}adlo, and Brodholt]{li2018elastic}
Yunguo Li, Lidunka Vo{\v{c}}adlo, and John~P Brodholt.
\newblock The elastic properties of hcp-fe alloys under the conditions of the
  earth's inner core.
\newblock \emph{Earth and Planetary Science Letters}, 493:\penalty0 118--127,
  2018.

\bibitem[Anderson(1963)]{anderson1963simplified}
Orson~L Anderson.
\newblock A simplified method for calculating the debye temperature from
  elastic constants.
\newblock \emph{Journal of Physics and Chemistry of Solids}, 24\penalty0
  (7):\penalty0 909--917, 1963.

\bibitem[Schreiber et~al.(1975)Schreiber, Anderson, Soga, and
  Bell]{schreiber1975elastic}
Edward Schreiber, Orson~L Anderson, Naohiro Soga, and James~F Bell.
\newblock Elastic constants and their measurement.
\newblock 1975.

\bibitem[Vo{\v{c}}adlo(2007)]{vovcadlo2007ab}
Lidunka Vo{\v{c}}adlo.
\newblock Ab initio calculations of the elasticity of iron and iron alloys at
  inner core conditions: Evidence for a partially molten inner core?
\newblock \emph{Earth and Planetary Science Letters}, 254\penalty0
  (1-2):\penalty0 227--232, 2007.

\bibitem[Kantor et~al.(2007)Kantor, Kantor, Kurnosov, Kuznetsov, Dubrovinskaia,
  Krisch, Bossak, Dmitriev, Urusov, and Dubrovinsky]{kantor2007sound}
Anastasia~P Kantor, Innokenty~Yu Kantor, Alexander~V Kurnosov, Alexei~Yu
  Kuznetsov, Natalia~A Dubrovinskaia, Michael Krisch, Alexei~A Bossak,
  Vladimir~P Dmitriev, Vadim~S Urusov, and Leonid~S Dubrovinsky.
\newblock Sound wave velocities of fcc {Fe--Ni} alloy at high pressure and
  temperature by mean of inelastic {X}-ray scattering.
\newblock \emph{Physics of the Earth and Planetary Interiors}, 164\penalty0
  (1-2):\penalty0 83--89, 2007.

\bibitem[Liu et~al.(2014)Liu, Lin, Alatas, and Bi]{liu2014sound}
Jin Liu, Jung-Fu Lin, Ahmet Alatas, and Wenli Bi.
\newblock Sound velocities of bcc-{Fe} and {Fe}$_{0.85}${Si}$_{0.15}$ alloy at
  high pressure and temperature.
\newblock \emph{Physics of the Earth and Planetary Interiors}, 233:\penalty0
  24--32, 2014.

\bibitem[He et~al.(2022)He, Sun, Kim, Jang, Li, and Mao]{he2022superionic}
Yu~He, Shichuan Sun, Duck~Young Kim, Bo~Gyu Jang, Heping Li, and Ho-kwang Mao.
\newblock Superionic iron alloys and their seismic velocities in earth’s
  inner core.
\newblock \emph{Nature}, 602\penalty0 (7896):\penalty0 258--262, 2022.

\bibitem[Yong et~al.(2019)Yong, Secco, Littleton, and Silber]{yong2019iron}
Wenjun Yong, Richard~A Secco, Joshua~AH Littleton, and Reynold~E Silber.
\newblock The iron invariance: implications for thermal convection in earth's
  core.
\newblock \emph{Geophysical Research Letters}, 46\penalty0 (20):\penalty0
  11065--11070, 2019.

\bibitem[Gomi et~al.(2013)Gomi, Ohta, Hirose, Labrosse, Caracas, Verstraete,
  and Hernlund]{gomi2013high}
Hitoshi Gomi, Kenji Ohta, Kei Hirose, Stephane Labrosse, Razvan Caracas,
  Matthieu~J Verstraete, and John~W Hernlund.
\newblock The high conductivity of iron and thermal evolution of the earth's
  core.
\newblock \emph{Physics of the Earth and Planetary Interiors}, 224:\penalty0
  88--103, 2013.

\bibitem[de~Koker et~al.(2012)de~Koker, Steinle-Neumann, and
  Vl{\v{c}}ek]{de2012electrical}
Nico de~Koker, Gerd Steinle-Neumann, and Vojt{\v{e}}ch Vl{\v{c}}ek.
\newblock Electrical resistivity and thermal conductivity of liquid {Fe} alloys
  at high {P} and {T}, and heat flux in earth's core.
\newblock \emph{Proceedings of the National Academy of Sciences}, 109\penalty0
  (11):\penalty0 4070--4073, 2012.

\bibitem[Anderson(1998)]{anderson1998gruneisen}
Orson~L Anderson.
\newblock The {G}r{\"u}neisen parameter for iron at outer core conditions and
  the resulting conductive heat and power in the core.
\newblock \emph{Physics of the Earth and Planetary Interiors}, 109\penalty0
  (3-4):\penalty0 179--197, 1998.

\bibitem[Wang et~al.(2010)Wang, Lv, Zhu, and Ma]{wang2010crystal}
Yanchao Wang, Jian Lv, Li~Zhu, and Yanming Ma.
\newblock Crystal structure prediction via particle-swarm optimization.
\newblock \emph{Physical Review B}, 82\penalty0 (9):\penalty0 094116, 2010.

\bibitem[Wang et~al.(2012)Wang, Lv, Zhu, and Ma]{wang2012calypso}
Yanchao Wang, Jian Lv, Li~Zhu, and Yanming Ma.
\newblock Calypso: A method for crystal structure prediction.
\newblock \emph{Computer Physics Communications}, 183\penalty0 (10):\penalty0
  2063--2070, 2012.

\bibitem[Baroni et~al.(2001)Baroni, De~Gironcoli, Dal~Corso, and
  Giannozzi]{baroni2001phonons}
Stefano Baroni, Stefano De~Gironcoli, Andrea Dal~Corso, and Paolo Giannozzi.
\newblock Phonons and related crystal properties from density-functional
  perturbation theory.
\newblock \emph{Reviews of Modern Physics}, 73\penalty0 (2):\penalty0 515,
  2001.

\bibitem[Kresse and Hafner(1993)]{kresse1993ab}
Georg Kresse and J{\"u}rgen Hafner.
\newblock Ab initio molecular dynamics for liquid metals.
\newblock \emph{Physical Review B}, 47\penalty0 (1):\penalty0 558, 1993.

\bibitem[Kresse and Joubert(1999)]{kresse1999ultrasoft}
Georg Kresse and Daniel Joubert.
\newblock From ultrasoft pseudopotentials to the projector augmented-wave
  method.
\newblock \emph{Physical Review B}, 59\penalty0 (3):\penalty0 1758, 1999.

\bibitem[Perdew et~al.(1996)Perdew, Burke, and
  Ernzerhof]{perdew1996generalized}
John~P Perdew, Kieron Burke, and Matthias Ernzerhof.
\newblock Generalized gradient approximation made simple.
\newblock \emph{Physical Review Letters}, 77\penalty0 (18):\penalty0 3865,
  1996.

\bibitem[Cococcioni and De~Gironcoli(2005)]{cococcioni2005linear}
Matteo Cococcioni and Stefano De~Gironcoli.
\newblock Linear response approach to the calculation of the effective
  interaction parameters in the {LDA+U} method.
\newblock \emph{Physical Review B}, 71\penalty0 (3):\penalty0 035105, 2005.

\bibitem[Majumdar et~al.(2017)Majumdar, John, Wu, and
  Yao]{majumdar2017superconductivity}
Arnab Majumdar, S~Tse John, Min Wu, and Yansun Yao.
\newblock Superconductivity in feh 5.
\newblock \emph{Physical Review B}, 96\penalty0 (20):\penalty0 201107, 2017.

\bibitem[Mao et~al.(1990)Mao, Wu, Chen, Shu, and Jephcoat]{mao1990static}
HK~Mao, Y~Wu, LC~Chen, JF~Shu, and Andrew~P Jephcoat.
\newblock Static compression of iron to 300 {GPa} and {Fe}$_{0.8}${N}i$_{0.2}$
  alloy to 260 {GPa}: Implications for composition of the core.
\newblock \emph{Journal of Geophysical Research: Solid Earth}, 95\penalty0
  (B13):\penalty0 21737--21742, 1990.

\bibitem[Giannozzi et~al.(2009)Giannozzi, Baroni, Bonini, Calandra, Car,
  Cavazzoni, Ceresoli, Chiarotti, Cococcioni, Dabo,
  et~al.]{giannozzi2009quantum}
Paolo Giannozzi, Stefano Baroni, Nicola Bonini, Matteo Calandra, Roberto Car,
  Carlo Cavazzoni, Davide Ceresoli, Guido~L Chiarotti, Matteo Cococcioni,
  Ismaila Dabo, et~al.
\newblock {QUANTUM ESPRESSO}: a modular and open-source software project for
  quantum simulations of materials.
\newblock \emph{Journal of Physics: Condensed Matter}, 21\penalty0
  (39):\penalty0 395502, 2009.

\bibitem[Li et~al.(2014)Li, Carrete, Katcho, and Mingo]{li2014shengbte}
Wu~Li, Jes{\'u}s Carrete, Nebil~A Katcho, and Natalio Mingo.
\newblock Shengbte: A solver of the boltzmann transport equation for phonons.
\newblock \emph{Computer Physics Communications}, 185\penalty0 (6):\penalty0
  1747--1758, 2014.

\end{thebibliography}

\end{document}